\documentclass{aa}

\usepackage{graphicx}
\usepackage{natbib,twoopt}
\usepackage{mathtools}
\bibpunct{(}{)}{;}{a}{}{,}
\usepackage[varg]{txfonts}
\usepackage{url}

\newcommand{\kms}{$\,$km$\,$s$^{-1}$}
\newcommand{\ergs}{$\,$erg$\,$s$^{-1}$}
\newcommand{\WHz}{$\,$W$\,$Hz$^{-1}$}
\def\HI{H{\,\small I}}
\newcommand{\msun}{${\mathrm{M_\odot}}$}

\newcommand{\mJybeam}{mJy beam$^{-1}$}

\begin{document}
\title{The LOFAR view of NGC~3998, a sputtering AGN}

\author{
Sarrvesh S. Sridhar\inst{1,2},
Raffaella Morganti\inst{1,2},
Kristina Nyland\inst{3},
Bradley S. Frank\inst{4,5},
Jeremy Harwood\inst{6},
Tom Oosterloo\inst{1,2}
}
\offprints 
{Raffaella Morganti, \\
\email{morganti@astron.nl}}

\authorrunning{Sridhar et al.}

\institute{Kapteyn Astronomical Institute, University of Groningen, P.O. Box 800, 9700 AV Groningen, The Netherlands
      \and ASTRON, the Netherlands Institute for Radio Astronomy, Postbus 2, 7990 AA, Dwingeloo, The Netherlands
      \and National Research Council, resident at the U.S. Naval Research Laboratory, 4555 Overlook Ave. SW, Washington, DC 20375, USA
      \and South African Radio Astronomical Observatory (SARAO), 2 Fir Street, Observatory, 7925, South Africa
      \and Department of Astronomy, University of Cape Town, Private Bag X3, Rondebosch 7701, South Africa
      \and Centre for Astrophysics Research, School of Physics, Astronomy and Mathematics, University of Hertfordshire, College Lane, Hatfield AL10 9AB, UK
\today
}


\abstract
{Low-power radio sources dominate the radio sky. They tend to be small in size and core-dominated, but the origin of their properties and the evolution of their radio plasma are not well constrained. Interestingly, there is mounting evidence that low-power radio sources can significantly impact their surrounding gaseous medium and, therefore, may be more relevant for galaxy evolution than previously thought. In this paper, we present low radio frequency observations obtained with LOFAR at 147~MHz of the radio source hosted by NGC~3998. This is a rare example of a low-power source which is extremely core-dominated, but which has two large-scale lobes of low surface brightness. We combine the new 147~MHz image with available 1400~MHz data to derive the spectral index over the source. Despite the low surface brightness, reminiscent of remnant structures, the lobes show an optically thin synchrotron spectral index~($\sim0.6$). We interpret this as being due to rapid decollimation of the jets close to the core, to high turbulence of the plasma flow, and entrainment of thermal gas. This could be the result of intermittent activity of the central AGN, or, more likely, temporary disruption of the jet due to the interaction of the jet with the rich circumnuclear ISM. Both would result in sputtering energy injection from the core which would keep the lobes fed, albeit at a low rate. We discuss these results in connection with the properties of low-power radio sources in general. Our findings show that amorphous, low surface brightness lobes should not be interpreted, by default, as remnant structures. Large, deep surveys (in particular the LOFAR 150~MHz LoTSS and the recently started 1400~MHz Apertif survey) will identify a growing number of objects similar to NGC~3998 where these ideas can be further tested.}


\maketitle

\section{Introduction}
One of the key ingredients all cosmological simulations agree on to be necessary for reproducing the observed properties of massive galaxies, is the (self-regulating) process which links the enormous amount of energy released by an active super massive black hole (i.e. active galactic nucleus, AGN) to the surrounding interstellar- (ISM) and intergalactic (IGM) medium (see \citealt{Heckman14} and references therein). Because of its importance, but also its complexity, a large effort (in observations and theory) is ongoing to constrain its properties.

In the past years, evidence has been mounting about the relevance of, among others, low-power radio sources for this process. 
Low-power radio sources ($\leq 10^{24}$ W/Hz) represent a class of relatively common AGN, being observed in at least 30\% of massive early-type galaxies (i.e. ${\rm M}_*> 10^{11}$ \msun, \citealt{mauch2007,best2005,Sabater19}). Their occurrence in massive galaxies is much higher than for  powerful radio AGN.

Among the relevant parameters defining the impact on the host galaxy is the duty-cycle of the activity (see e.g. \citealt{Ciotti10,Gaspari13,Morganti17}). Interestingly, low-power radio sources hosted by massive elliptical galaxies have been suggested to be {\sl `on'} most of the time with only short intervals of no activity, therefore almost continuously dumping energy on the host galaxy (\citealt{Sabater19}). This, however, has not yet been confirmed by detailed studies of radio AGN (see e.g. \citealt{Nyland16}).
Finally, a growing number of observations have shown that indeed the gas in the central regions can be affected also by low power jets (\citealt{Croston08,Combes13,Morganti15,Rodriguez17,Fabbiano18,Husemann19} and refs therein) and numerical simulations predict this should occur \citep{Mukherjee18a,Mukherjee18b}. 

It is therefore important to consider the effect of this class of objects, which represents the most common population of radio AGN in the nearby Universe. 

The morphology and properties of the radio emission appears to change with decreasing radio powers. Moving to low radio power (below $\sim 10^{23} - 10^{24}$ \WHz), the radio emission is often compact (e.g., a few kpc or less) compared to the extent of the host galaxy, and more core dominated (see e.g. \citealt{Balmaverde06,Baldi09,Baldi19}). The radio morphology of these objects could be the result of different physical properties or different evolutionary paths. 
For example, low power sources tend to have low velocity jets and, as a result of this, their jets are characterised by a larger component of turbulence compared to jets in powerful radio sources. This has been derived from detailed modelling of jets in  Fanaroff-Riley type I (FRI, \citealt{Fanaroff74}) radio galaxies (see e.g. \citealt{laing2014,Massaglia16}) and these characteristics are expected to be even more extreme in lower-power jets, which have lower bulk Lorentz factors \citep{Baldi19}. The turbulence and the consequent entrainment of the external thermal medium, gives rise to high thermal fraction inside these jets, making them slow  further  down  (e.g. \cite{Bicknell98}).

Low-luminosity radio AGN have been suggested to be fuelled via a variety of mechanisms (see \citealt{Storchi19} for a review). This includes hot gas in advection dominated accretion flows (\citealt{Yuan14}, and references therein), chaotic cold accretion (\citealt{Gaspari13,Gaspari15} from gas cooling from the hot halo (e.g. \citealt{Negri15}), secular processes. The availability of this gas  as fuel allows the nuclear activity to be restarted after only short interruption, thus supporting their recurrent nature with a high duty cycle.

Because of these physical properties and their galactic-scale jets, low-luminosity radio galaxies are also particularly sensitive to the external conditions. Numerical simulations show that the coupling between energy released by the jet and the ISM remains high also for low power jets (e.g. \citealt{Mukherjee18a,Mukherjee18b}). Because less efficient in accelerating clouds, these jets are trapped in the ISM for a longer period of time, thereby increasing the impact they can have on the ISM. 
This further suggests that these objects could play a role in the evolution of the host galaxy. Because of this, it is important to understand more about their life and the properties of the their jets and how they evolve.

We address this by studying one of these objects, the radio source hosted by NGC~3998\footnote{In this paper, we will refer to it by the name of its optical identification}, which provides a unique case-study.
NGC~3998 represents one of the rare examples of a low-power radio AGN $(\mathrm{P}_\mathrm{1400~MHz}\sim 10^{22} \mathrm{W~Hz}^{-1})$ with extended radio emission ($\sim 20$ kpc)\footnote{In this paper, we assume a distance of of 13.7 Mpc ($z=0.0035$) for NGC~3998 \citep{Cappellari11}. At this distance, 1\arcsec\ corresponds to 65.9 pc.}. 
Here we expand the study presented in \citet{frank2016} and focus on the morphology at low frequency (147~MHz) and, in combination with the previously published 1400~MHz image, the radio spectral index.  This  information allows us to place new constraints on the age and evolution of the radio emission.

The paper is structured in the following way. In Sec. \ref{sec:target} we describe the properties of the target of our case-study, NGC~3998 and its associated radio source. In Sec.\ref{sec:LOFARobservations} we present our new LOFAR observations. We discuss the results including the spectral index distribution in Sec. \ref{sec:results}. Section \ref{sec:nature} is dedicated at discussing the nature of the radio emission and the implications for other groups of low-power radio sources. We present our conclusions in Sec. \ref{sec:conclusions}.

\section{The target: NGC~3998}
\label{sec:target}

\begin{figure}
\resizebox{\hsize}{!}{\includegraphics[clip=True,trim={1.3cm 18.75cm 11cm 1.0cm}]{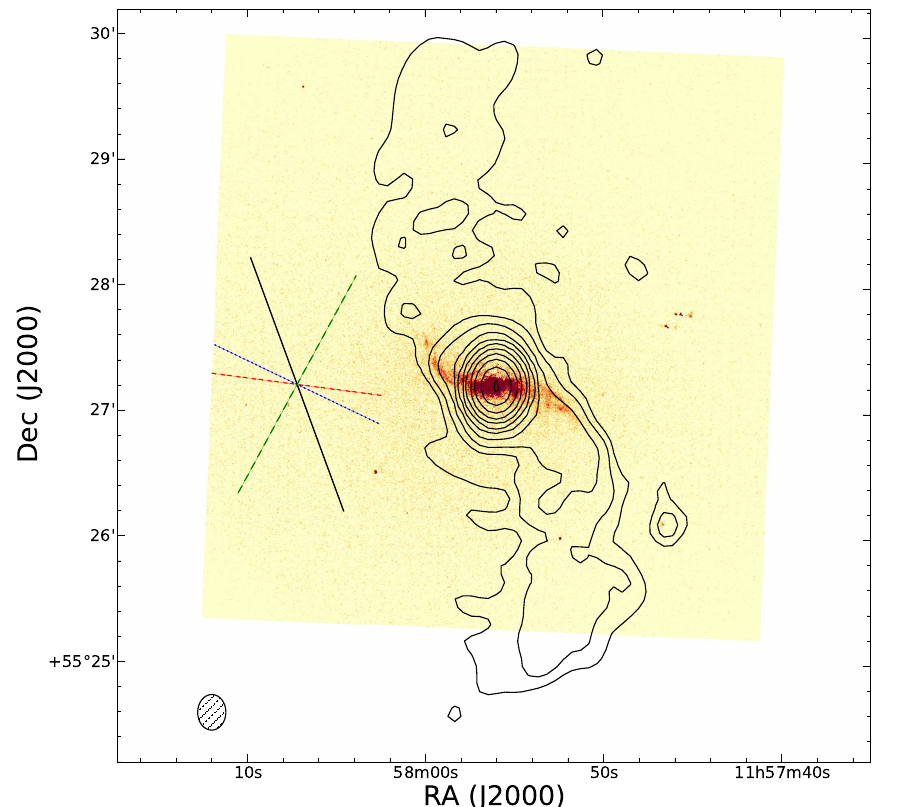}}
\caption{Radio continuum contours from the narrow-band 21-cm observations of NGC~3998 using the Westerbork Synthesis Radio Telescope overlaid on the continuum-subtracted H$\alpha$ image from \cite{sanchez-gallego2012}. On the left, we show a depiction of the position angles of the various galaxy components: green-dashed: stars (PA$\sim 135\degr$ from \citealt{krajnovic2011}), black-solid: inner $\sim 5$ kpc of radio continuum (PA $\sim 33\degr$), blue-dotted: \HI\ disk (PA $\sim 65\degr$), red-dot-dash: H$\alpha$ disk (PA$\sim 84\degr$). Image taken from \cite{frank2016}.}
\label{fig:franketal}
\end{figure}

As described in \cite{frank2016}, the radio source in NGC~3998 shows two S-shaped lobes (see Fig. \ref{fig:franketal}) of low surface-brightness ($\sim 5$ mJy/arcmin$^2$ at 1400~MHz). In addition, a bright central component is also observed.  
Because of this, the radio source is characterised by an extremely high core dominance (S$_\mathrm{core}$/S$_\mathrm{ext} = 8$ or S$_\mathrm{core}$/S$_\mathrm{tot} = 0.9$, derived at 1400~MHz, see \citealt{frank2016}), much higher than  typically found for FRIs \citep{morganti1997,laing2014} and in the radio sources of lower power studied by \cite{baldi2015}. 
From a morphological point of view, the lobes of NGC~3998 have characteristics which may suggest that they are still actively being fed by the nucleus (see \citealt{frank2016}). However, low surface-brightness structures have been often suggested to be associated with remnants, not fed anymore by active jets (e.g. \citealt{Saripalli12,Brienza16}).  

NGC~3998 is part of the ATLAS$^{3\rm D}$ sample \citep{Cappellari11} of nearby early-type galaxies. The radio study of this sample (e.g. \citealt{Serra12,Nyland16,Nyland17}) indicated that the type of radio structure found in NGC~3998 tends to be rare in low power radio sources.  
Only three objects out of more than 100 for which deep continuum images are available (either at low or high resolution; \citealt{Serra12,Nyland16}) were found to have extended emission on the scale of tens of kpc, one of them being NGC~3998 (the other two NGC~3665 and NGC~5322). 

NGC~3998 is a gas-rich early-type galaxy. Atomic neutral hydrogen was observed in this galaxy  distributed on a nearly polar disc structure extended over 10 Kpc and with a mass of M$_{\HI} =  4\times 10^8$ \msun)  (\citealt{Serra12, frank2016}). 
In ionised gas, Devereux (2018)
report the detection of broad permitted and forbidden emission lines using the high spatial resolution of STIS-HST. They suggest the presence of kinematically disturbed gas under the effect of the emission of the AGN.
On larger scales, the S-shaped radio lobes mirror the warp of the gas (\HI\ and H$\alpha$) as illustrated in Fig. \ref{fig:franketal}. This has brought the suggestion that, in the inner kpc regions of the galaxy, the stellar torques may be causing the gas disk to warp into an S-shape. These torques can also produce orbit crossings and shocks in the gas, which causes the gas to lose angular momentum, accreting onto a central black hole and fueling the AGN. 
Thus, the fuelling of the central super-massive black hole could be occurring through accretion of cold gas via “discrete events”, as also suggested by the observed variability of the radio core \citep{frank2016}. However, optically thin hot accretion flow is also a possibility as shown by the {\it NuSTAR} and {\it XMM-Newton} X-ray data (\citealt{Younes19}).

All of these properties make NGC~3998 particularly suitable to use as a case-study for understanding the evolution of the radio plasma in low-power AGN.
One way to explore the evolution of these sources is to assess the spectral properties of their radio lobes. We perform this type of analysis for NGC~3998 using low-frequency observations at 147 MHz carried out with the LOw Frequency ARray (LOFAR, \citealt{Haarlem13}) combined with available data at 1400~MHz presented in \cite{frank2016}.

\section{LOFAR observation and data reduction}
\label{sec:LOFARobservations}

NGC 3998 was observed for eight hours with the LOFAR High Band Antenna (HBA) on March 23, 2015 (Open Time project LC3\_025 PI K. Nyland). The eight hour scan on NGC 3998 was bracketed with two 15-minute scans on the flux density calibrator 3C~295. An overview of the observational parameters are presented in Table \ref{tab:obsParam}. The target and the calibrator were observed with an identical frequency setup covering a bandwidth ranging from 123.7 MHz to 172.2 MHz. The frequency range was divided into 248 195.3125 kHz wide subbands (SBs) which were further sub-divided into 64 channels each.

After the observation, the raw visibilities were first flagged for radio frequency interference (RFI) before averaging down to a frequency resolution of 4 channels per sub-band and 2s time resolution. RFI flagging and averaging were carried out within the standard LOFAR pre-processing pipeline framework \citep{heald2010} which uses AOFlagger \citep{offringa2010, offringa2012} for RFI-flagging. 
Due to the large size of the raw dataset, only the averaged data products were uploaded to the LOFAR Long Term Archive (LTA)\footnote{\url{https://lta.lofar.eu}}. After pre-processing, the data were calibrated and imaged using the facet calibration scheme described in \cite{vanWeeren2016} and \cite{williams2016} which carries out calibration in two stages: a direction independent step followed by a direction dependent calibration procedure.

In the direction independent step, we first removed bad stations (CS013HBA0 and CS013HBA1) before averaging the calibrator measurement sets to 1 ch/sb and 8s time resolution. Using the ``Black Board Self-calibration'' software package (BBS, \citealt{pandey2009}), we obtained amplitude and phase solutions for the RR and LL correlations using the 3C~295 model assuming the flux scale defined in \cite{scaife2012}. From the gain solutions, we computed the clock offset and gain amplitude for each LOFAR station and transferred these to the target measurement sets using BBS. Phase calibration was then applied to the clock and gain amplitude corrected target data using a $6~\times~6$ deg$^2$ model of the sky from the TIFR GMRT Sky Survey \citep[TGSS~ADR1,][]{intema2016}. The extracted skymodel contains 1504 point and 1144 Gaussian components.

The aim of the direction dependent calibration procedure is to minimize the artifacts produced by the time and position dependent errors due to the ionosphere and the LOFAR station beam. To achieve this, we chose 15 directions or facets such that each facet contains at least one point source brighter than 750 mJy. For each facet, we performed self calibration to derive good calibration solutions and models of the sources in that direction. Using the new calibration solutions, we then subtracted out the sources in each of the 15 facets. Finally, the target facet containing NGC 3998 was corrected with calibration solutions derived using the bright source VLSS J1156.5+5520 which is $\sim13\arcmin$ away from NGC 3998. 

The direction-dependent calibrated visibilities were imaged using the wideband deconvolution algorithm available in WSClean \citep{offringa2014} imager. We produced images of NGC~3998 at two different angular resolutions. The high resolution image was produced by weighting the visibilities using the Briggs weighting scheme with robust$=-0.5$ \citep{briggs1995} and applying a $5\arcsec$ Gaussian taper. The restored image has an angular resolution of $11\farcs6 \times 7\farcs7$ and the rms noise close to NGC~3998 is about $170\,\mu$Jy/beam. To study the low surface brightness emission from the radio lobes, we also produced a low resolution image using robust $=-0.7$ and applying a $25\arcsec$ Gaussian taper. The restored image was convolved with a $30\arcsec$ circular Gaussian beam and the rms noise in the image is $680\,\mu$Jy beam$^{-1}$.

We compared the integrated flux densities of background sources within our field of view with the flux densities reported in the TGSS ADR1 data release. We do not find any systematic offset in the integrated flux densities. We assume a 15\% uncertainty for all the 147~MHz flux densities reported below. 

\begin{table}
\caption{LOFAR HBA Observational parameters}
\label{tab:obsParam}
\begin{tabular}{ll}
\hline \hline
Parameter               & Value \\
\hline
Project ID              & LC3\_025 \\
Observation ID          & 294275 (3C~295) \\
                        & 294271 (NGC~3998) \\
Pointing centres        & 14:11:20.5 +52:12:09.9 (3C~295) \\
                        & 11:57:56.1 +55:27:12.9 (NGC~3998)\\
Total on-source time    & 8.0 hr \\
Observation date        & 2015 March 23 \\
Frequency range         &  123.7~--~172.2 MHz\\
SBs                     & 248 SBs \\
Bandwidth per SB        & 195.3125 KHz \\
LOFAR Array Mode        & HBA Dual Inner \\
Stations                & 61 total \\
\hline
\end{tabular}
\end{table}

One of the goals of this work is to derive the spectral index of the various structures of the source, i.e. nuclear region and lobes. We have done this by using both the integrated flux densities and by constructing a spectral index image using the LOFAR and the WSRT from \citet{frank2016}. The spectral index $\alpha$ is defined as $S \sim \nu^{-\alpha}$. The integrated flux densities are derived over boxes including these structures and summarised in Table \ref{tab:fluxes}. 

\section{Results}
\label{sec:results}

\subsection{NGC~3998 at low radio frequency}

\begin{figure}
\resizebox{\hsize}{!}{\includegraphics[width=8cm,angle=-90]{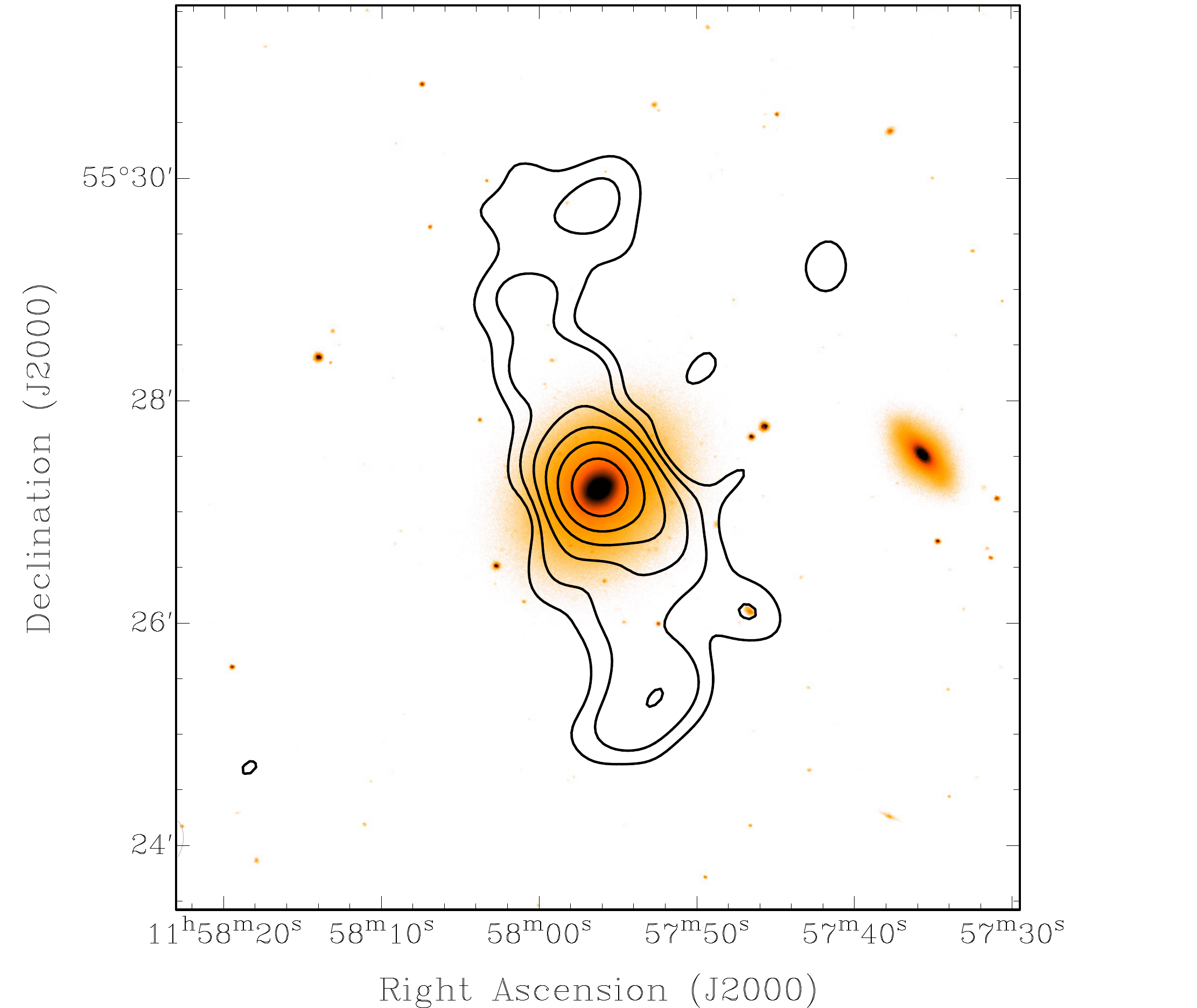}}
\caption{Overlay of the $30\arcsec$ resolution Stokes I image of NGC~3998 (contours) and an $r$-band optical image from the Sloan Digital Sky Survey \citep{alam2015}. The LOFAR contours are drawn at -1.8, 1.8 to 94 \mJybeam in step of 2. 
}
\label{fig:totInt30asec}
\end{figure}

Figure~\ref{fig:totInt30asec} shows the low-resolution image of NGC~3998 where the LOFAR total intensity contours are overlayed on an optical $r$-band image from the Sloan Digital Sky Survey \citep{alam2015}. Our LOFAR image reveals the same S-shaped radio lobes seen in the WSRT 1400~MHz radio images from \cite{frank2016}. The LOFAR image at higher spatial resolution (although still affected by artifacts, which are likely due to unmodelled ionospheric effects) allows us to better explore the nuclear regions. Figure \ref{fig:totInt7asec} shows the high resolution image of NGC~3998 superposed onto the low resolution image.

The high-resolution image shows that the inner structure has, in addition to the bright core, a small extension to the N and a much clearer extension in the SW direction. The latter follows nicely the structure seen in the WSRT  image at 1400~MHz.
 Interestingly, on the VLBI scale the source shows a very small - about 1~pc - one-sided jet elongated in the Northern direction \citep{Helmboldt07},  shown in the inset of Fig. \ref{fig:totInt7asec}. 
Thus, the jets in NGC~3998 appears to change direction while expanding in the inner kpc regions. This is possibly due to precession, perhaps connected to the warping structure seen inside the inner regions (Frank et al. 2016). It is, however, difficult to establish with the available data whether this change in the direction of the jet can be induced by interaction with the circumnuclear medium. At larger distances the jet-lobes continues to warp, producing the S-shape structure.

The high core dominance, although not as extreme as at 1400 MHz, is also confirmed in the LOFAR image at 147~MHz where we derive $S_\mathrm{core}/S_\mathrm{tot} = 0.32$, 
 using the peak of the emission in the high resolution image as core flux density. 

\subsection{Spectral index distribution}

The spectral index image was constructed, following the procedure described in \cite{Harwood13}, using the 147 MHz and the WSRT 1400~MHz, with matched uvrange $(0.28 - 13.3~\mathrm{k}\lambda)$. The radio continuum images were convolved to a common 
resolution of $30\arcsec$ and placed on the same coordinate grid. 
All pixels below $5 \sigma_\mathrm{rms}$ were masked and the spectral index error per pixel was estimated using the relation:

\begin{equation}
\alpha_{err} = \frac{1}{\log \frac{\nu_\mathrm{lofar}}{\nu_\mathrm{wsrt}}} \sqrt{\left( \frac{S_\mathrm{lofar,err}}{S_\mathrm{lofar}} \right)^2 + \left( \frac{S_\mathrm{wsrt,err}}{S_\mathrm{wsrt}} \right)^2}
\end{equation}
where $S_\mathrm{lofar}$ and $S_\mathrm{wsrt}$ are the pixel values in the LOFAR and the WSRT maps, and $S_\mathrm{lofar,err}$ and $S_\mathrm{wsrt,err}$ are the uncertainties on each pixel value. Flux density calibration errors of 5\% and 15\% were assumed for WSRT and LOFAR respectively.  

Given the variability of the nuclear regions as discussed in \citep{frank2016}, using our observations to derive the spectral index of the nuclear region needs to be done with care. In principle simultaneously, or observations at least close in time are needed. 

Figure \ref{fig:specIndex} shows the distribution of the spectral index (left) across the radio source between 147 MHz and 1400~MHz and the corresponding errors (right). The main result is that the low-surface brightness lobes have a spectral index typical of active lobes of radio galaxies with an average value of $\alpha^{\rm 1400 MHz}_{\rm 147 MHz} \sim 0.6$. 
As seen in Fig. \ref{fig:specIndex}, the spectral index ranges between 0.4 and 0.8. Furthermore, we do not see any trend in the spectral index moving away from the core. A trend (i.e. steepening of the spectral index with the distance from the core) is often been seen in FRI with tailed structures as described, for example, in \cite{Parma02}. This suggests that in NGC~3998 the conditions are different from those observed in radio galaxies where the spectral shape is dominated by the energy losses due to synchrotron emission. The presence of turbulence inducing a strong mixing of the electron populations could be at the origin of the relatively homogeneous spectral index seen in the lobes of NGC~3998 (see Sec. \ref{sec:nature}). 
An inverted spectrum ($\alpha^{\rm 1400 MHz}_{\rm 147 MHz} \sim  - 0.23\pm0.10$) is instead clearly observed in the central region. 

The spectral indices of the lobes and of the nuclear region are also confirmed by the results from the integrated spectral indices (see Table \ref{tab:fluxes}). The integrated flux densities of the lobes are derived from rectangular regions excluding the central component while  
for the central region, the spectral indices could be derived from the higher resolution images and in two frequency ranges. 
For the central region, an inverted spectral index is confirmed at low frequencies ($\alpha^{\rm 1400 MHz}_{\rm 147 MHz} \sim  - 0.35\pm0.02$), turning over at higher frequencies ($\alpha^{\rm 4800 MHz}_{\rm 1400 MHz} \sim  +0.18\pm0.12$), see Table \ref{tab:fluxes}. 
The flatter spectral index obtained at low frequencies compared to what see in the image of Fig. \ref{fig:specIndex}, is likely due to the lower spatial resolution of the latter (i.e. where the peak flux density at the core location includes a larger fraction from the beginning of the jet). 

As mentioned above, we need to be careful about the interpretation of the exact value of the spectral index in the central regions given the variability noted in the flux density and the observations not being simultaneous (although quite close in time, see Notes in Table \ref{tab:fluxes}). The flux densities  
obtained from the 1400 MHz and 4900 MHz were taken only three months apart, i.e. the closest we have in time (see Fig. 3 in \citealt{frank2016}). 

That notwithstanding, the overall spectrum of the nuclear regions seems to be consistent with a flat or self-absorbed spectrum, as expected if the central region is dominated by a very small source, $< 2\arcsec$ (100 pc, as suggested by the work of \citealt{Helmboldt07}). 
However, it is interesting to note that the slopes derived are not particularly steep and the spectral shape of the nuclear regions appears to be similar to what derived for  FR0 radio sources (see Capetti et al. submitted).
The self-absorbed and bright core suggests that the core is the base of a relativistic synchrotron jet seen in powerful young radio galaxies.

\begin{table}
\centering
\caption{Integrated flux densities of NGC~3998 taken from this paper and from \cite{frank2016}.}
\label{tab:fluxes}
\begin{tabular}{lrr}
\hline\hline
Freq & Resolution &Flux density  \\
 MHz & arcsec &mJy  \\
\hline\hline
{\bf Core}\tablefootmark{a} &  & \\
\hline
147~MHz & 5  & 68$\pm$10  \\
147~MHz & 30 & 94$\pm$14 \\
1400~MHz & 15 & 149$\pm$7 \\
4901~MHz & 5 & 118$\pm$6  \\
\hline
{\bf Lobes}\tablefootmark{b}  \\
\hline
147~MHz  N & 30 & $31.3\pm7.6$  \\
147~MHz S & 30 & $41.4\pm8.6$  \\
1400~MHz  N & 30 & $8.0\pm1.1$ \\
1400~MHz  S & 30 & $12.2\pm1.4$  \\
\hline
\end{tabular}
\tablefoot{
\tablefoottext{a}{Peak flux density. The 147, 1381, and 4901~MHz flux densities were recorded on 23/03/2015, 02/06/2015, and 20/06/2015 respectively.}\\
\tablefoottext{b}{Integrated on the same box - excluding the central region - at the two frequencies.}
}
\end{table}

\begin{table}
\centering
\caption{Integrated spectral indices of NGC~3998 derived from the flux densities of Table \ref{tab:fluxes}. }
\label{tab:spectralindex}
\begin{tabular}{lrrl}
\hline\hline
 Region  &  \multicolumn{2}{c}{$\alpha$} \\
         & $_{\rm 147 MHz}^{\rm 1400 MHz}$  & $_{\rm 1400 MHz}^{\rm 4900 MHz}$ \\
\hline\hline
{\bf Core} &  $-0.35\pm0.02$  &  $+0.18\pm0.12$   \\
{\bf N lobe} & $+0.60\pm0.07$ & \\
{\bf S lobe} & $+0.54\pm0.05$ & \\
\hline
\end{tabular}
\end{table}

\section{The nature of the radio emission in NGC~3998}
\label{sec:nature}

\begin{figure}
{\includegraphics[width=10cm,angle=0]{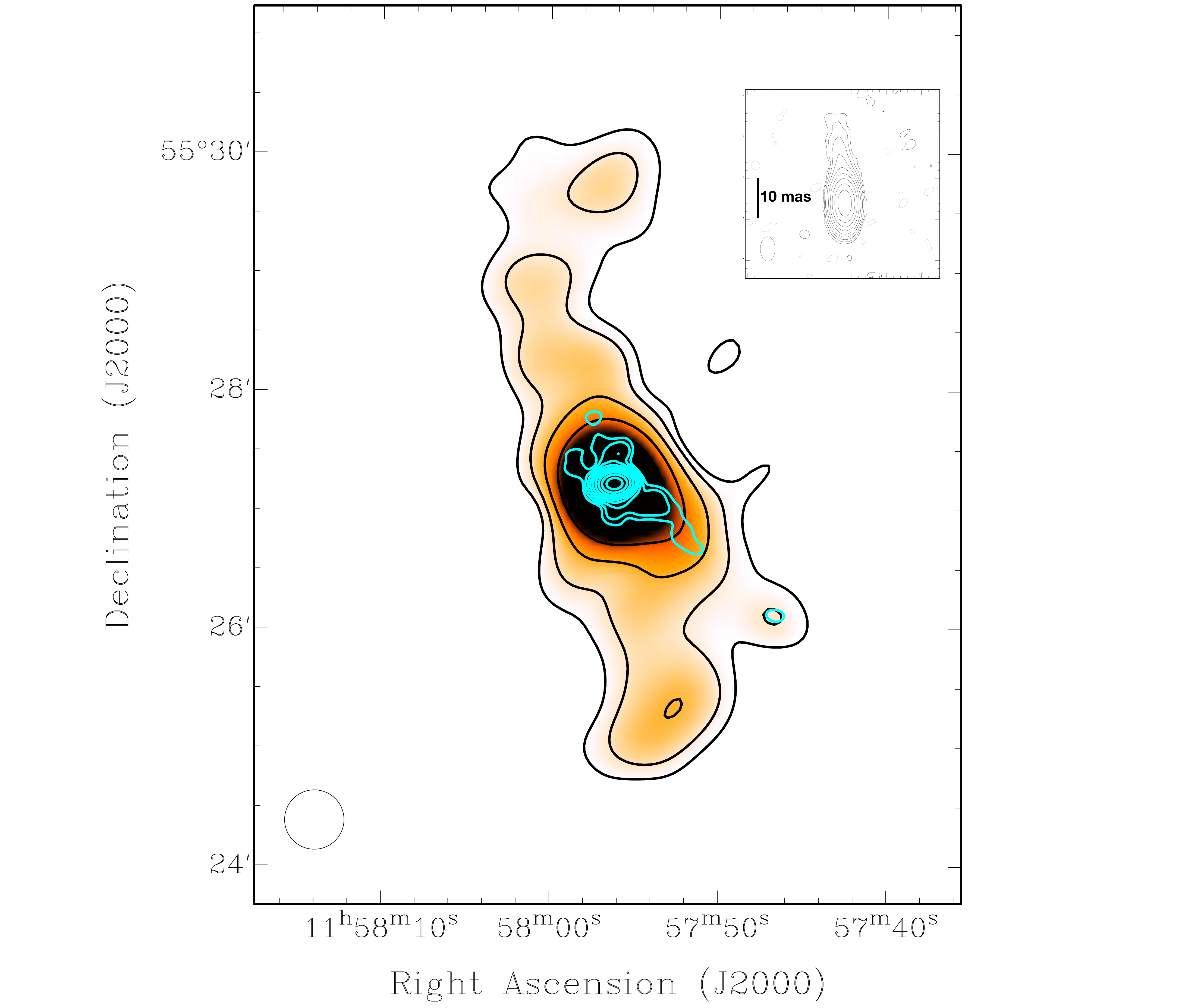}}
\caption{LOFAR image (30\arcsec resolution) with superposed the contours of the high resolution LOFAR image. Contour levels are from -1.8 1.8 to 94 \mJybeam in multiple of 2 and -1.8, 1.8 to 68 \mJybeam\ in steps of 1.5  respectively. The inset on the top-right corner represents the VLBA image from \cite{Helmboldt07} with contours level -0.5 0.5 to 200 \mJybeam\ in step of 2 . The size of the VLBA jet is about 15 mas, corresponding to about 1 pc.}
\label{fig:totInt7asec}
\end{figure}

\begin{figure*}
\centering
{\includegraphics[width=8cm,angle=0]{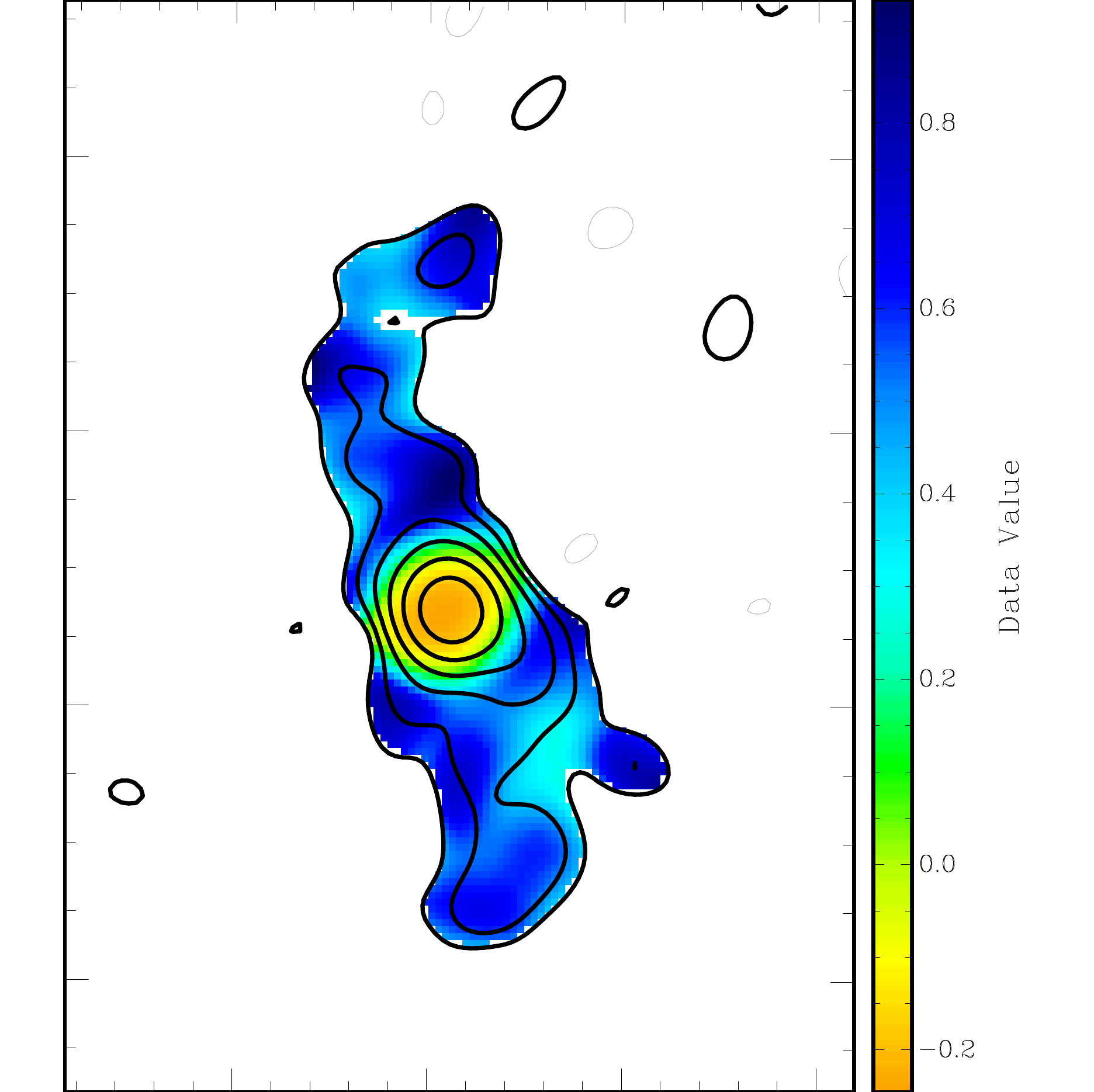}
\includegraphics[width=8cm,angle=0]{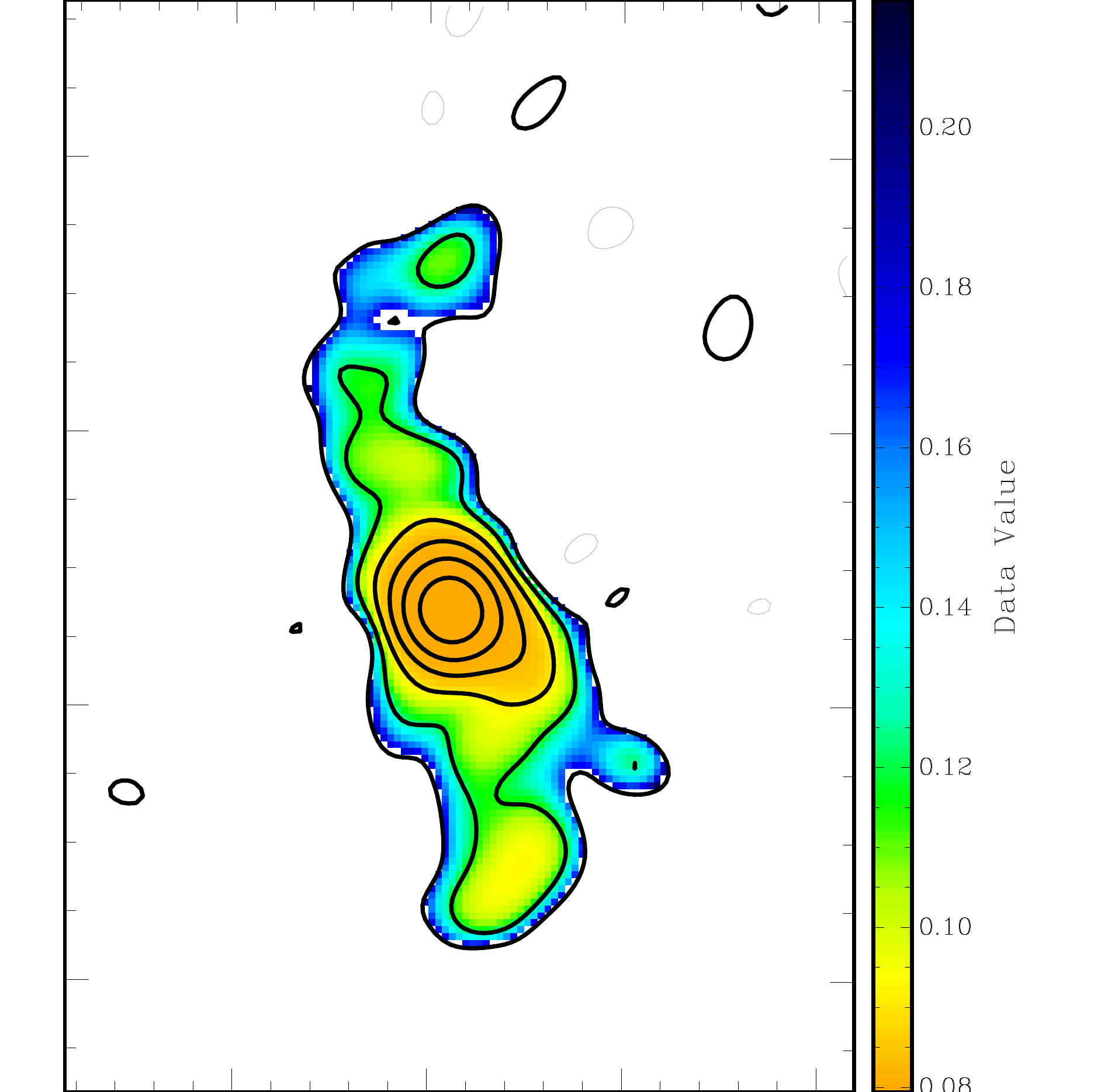}}
\caption{Spectral index image {\sl (left)} and associated error {\sl (right)} between 147 MHz and 1400 MHz, with superimposed contours as in Fig. \ref{fig:totInt30asec}.}
\label{fig:specIndex}
\end{figure*}

The main characteristics of the radio emission in NGC~3998 can be summarised as following. 
The 147~MHz LOFAR image shows two low-surface brightness lobes ($\sim 13$ mJy arcmin$^{-2}$ at 147~MHz) with similar extent as was seen in the 1400~MHz radio image by \cite{frank2016}.
The lobes are reminiscent of plumes seen in FRI-tailed radio galaxies (although typically at larger distances from the core) . 
They also show an S-morphology that appears to follow the warped structure of the inner gas. This could also be reminiscent of cases of precessing jets (e.g. \citealt{Kharb06,Kharb10}).
The high-resolution LOFAR image traces the starting of the jets in the nuclear regions (see Fig. \ref{fig:totInt7asec}). The dominance of the nuclear region compared to the extended lobes is confirmed at low frequencies.

It is interesting to note that, despite the low surface brightness, the spectral analysis suggests that the lobes are not old remnant structures. The integrated spectral indices of the lobes between 147 MHz and 1400 MHz are found to be around $\alpha^{\rm 1400 MHz}_{\rm 147 MHz} \sim 0.6$ (see Fig. \ref{fig:specIndex} and Table \ref{tab:fluxes}), consistent with values expected at these frequencies in active radio sources (see e.g. \citealt{Heesen18}). These values are also consistent with no major steepening occurring at frequencies lower than 1400~MHz. Furthermore, the spectral indices are relatively uniform across the lobes (see Fig. \ref{fig:specIndex}), with no trend, i.e. steepening,  with the distance from the core.  

We estimated that the magnetic field strength in NGC~3998 to be 1~nT ($10\mu$G). We calculated the magnetic field strength in NGC~3998 following the equation from \cite{Worrall2006} assuming that the magnetic field and the particles satisfy energy equipartition. We also assumed a power law distribution for the particles of the form $\mathrm{N}(\gamma) \propto \gamma^{-\alpha}$ with a minimum and maximum Lorentz factors of $\gamma_\mathrm{min}=10$ and $\gamma_\mathrm{max}=10^6$ respectively. Following \cite{Brienza16}, we estimated that the source volume to be $3.5 \times 10^{66}~\mathrm{cm}^3$ assuming a prolate ellipsoidal geometry with major and minor axes equal to the projected size of the source (i.~e. $90\arcsec$ and $60\arcsec$ ) respectively. We further assumed that the particle energy content of the source to be equally distributed between heavy-particles and electrons. 

With a magnetic field strength of 1~nT (10 $\mu$G), a break frequency of 1400~MHz would provide an upper limit on the age of the radio lobes of 38 Myr at the redshift of NGC~3998. 
A note of caution is needed, given that the spectral index is derived only for low radio frequencies, i.e. below 1400~MHz. Our analysis still leaves the possibility that the emission is associated with remnant of a recently switched off source where the  break frequency, $\nu_\mathrm{off}$, has not reached the low frequencies yet. Because of this, the age quoted above could be considered as an upper limit.

This age can be compared with what was obtained from the analysis presented in \cite{frank2016}. There, it was assumed an average velocity of propagation of the jets $\sim 1700$ \kms\ (following the results for another low-power radio galaxy, Centaurus~A, by \citealt{Croston09}) and derived time scales of a few times $10^6$ to at most $10^7$ yr (see \citealt{frank2016} for the full discussion). These times are shorter than what derived from the consideration on the spectral properties (that indeed was an upper limit) but consistent with no steepening be present at frequencies lower than 1400~MHz. 

The spectral index derived for the lobes, the S-shape of the lobes and its similarity with the inner warp and the presence of jet-like structures observed in the high-resolution LOFAR image (see Fig. \ref{fig:totInt7asec}), suggest the scenario where the lobes are still actively fed by the central region - or where the injection is only very recently stopped. However, more is required to fully explain the  morphology of the source, e.g. the bright nuclear region and the extreme low-surface brightness of the lobes, and what this tells us about the evolution of the radio source. 
 
\subsection{A ``sputtering'' radio source?}
\label{sec:sputtering} 

 Two possible scenarios can be considered to describe all the properties of the radio emission in NGC~3998: i) although we have excluded the lobes to be {\sl old} remnants, the overall radio structure could still be the result of {\sl intermittent fuelling}; ii) alternatively, it could be the result of {\sl intermittent flow} due to a strong interaction between a recently created jet and the surrounding rich medium, with the subsequent temporarily decollimation of the flow. The two scenarios are not necessarily disconnected and, actually, can be complementary. 

The idea that the activity is intermittent is supported by the variability of the core of NGC~3998 pointed out by a number of authors (e.g. \citealt{Kharb12}, \citealt{frank2016}),
suggesting that the fuelling may happen via discrete events.
From the morphology and the spectral information, the time-scales of the intermittent activity can be - at least to first order - derived.
One cycle of activity should have started very recently as shown by the presence of the relatively bright core from which jets are emerging (see e.g. \citealt{Helmboldt07} ad Fig. \ref{fig:totInt7asec}). The previous cycle is traced by the low-surface brightness lobes and it happened, at most, a few tens of Myr ago as discussed above.
The adiabatic expansion of the lobes could be responsible for the low surface brightness of these structures. However, the lack of any gradient in the spectral index inside the lobe would require e.g. the presence of turbulence capable, due to diffuse shock acceleration (i.e. second order Fermi acceleration), of reaccelerating old electrons but also to provide  population mixing that would dilute any spectral gradient. This would help preventing the steepening of the spectral index due to radiative (and adiabatic) losses. This possibility has been discussed also in the case of other radio galaxies (e.g. \citealt{Brienza18}, see also below). This process does not last more than a few tens of Myr after which the lobes quickly fade away (\citealt{Eilek14}). 
On the other hand, the time-scales of the intermittent activity, i.e. the {\sl off time}, should not be too long otherwise the spectrum would be steeper than what observed. 
Thus, in this first scenario, the origin of the high contrast between the core and the lobes would be due to the intermittent nature. However, the origin of the high turbulence that would dilute any spectral gradient is less clear.

The second scenario - strong interaction able to {\sl temporarily} disrupt the jet - is motivated by the abrupt change in properties of the jets (transitioning to poorly collimated, low surface-brightness plumes) as soon as they are outside the nuclear regions and by the fact that the radio source resides in a gas rich galaxy.

From the \HI\ observations presented in \cite{frank2016} we could not derive any sign of interaction between the jet and the gas. However, more recent observations of the ionised gas presented in \cite{Devereux18} may hold clues. 
\cite{Devereux18} report the detection of broad permitted and forbidden emission lines using the high spatial resolution of STIS-HST. They rule out the radiatively driven outflow scenario to explain the broad forbidden lines (e.g. [OIII] with a FWHM of $2185 \pm 265$ \kms)  because the AGN is simply unable to provide sufficient radiation pressure. The possibility of the jet to provide the energetics is ruled out by the author because the jet direction is perpendicular to the position angle of the STIS slit. However, at least the first part of the jet (e.g. the VLBI jet) would be included in the slit (which has a width of 0.1\arcsec) and its power can, in principle, provide the energetics to produce the outflow. Furthermore, the region affected by the jet could be larger than the jet itself, for example a cocoon, see  simulations by \cite{Mukherjee18a}. If this is the case, this may support the presence of a strong interaction between jet and ISM in the nuclear regions.
In \cite{frank2016} we have estimated a jet power of $P_{\rm jet} \sim 1.64\times10^{42}$ \ergs\ derived using the relation given in \cite{Cavagnolo10}. This is likely to be a lower limit to the jet power, given the likely additional presence of a substantial thermal component (see e.g. the case of IC~5063, \citealt{Mukherjee18a}). Under these circumstances, the radio plasma would be able to drive the outflow.

Thus, in this second scenario, the low-power jet is not necessarily intermittent but is undergoing a strong interaction with the ISM. Because of this, the jet is temporarily trapped (and perhaps temporarily disrupted), piling up energy in the inner region of the galaxy, until bursting and expanding in the extended lobes. This interaction would decollimate the flow and, combined with the low velocity of the jet, would result in a {\sl turbulent flow}, with an increased thermal component due to the entrainment of material from the surrounding medium. This would further slow down and diffuse the jet and make the lobes to evolve into low surface brightness structures. 

Interestingly, in NGC~3998 the lobes appear to have kept the imprint of the warp seen in the central sub-kpc regions. This may further suggest that the lobes are still connected (and fed, even if at low rate) to the central region. This has been found to be the case in Centaurus A (see below), but high sensitivity combined with high spatial resolution will be needed to investigate the possible presence of a connecting channel in NGC~3998. 


Thus, the morphology and characteristics of the radio source in NGC~3998 can be broadly described as resulting from intermittent emission from the central regions. This intermittency can be caused by genuine intermittent fuelling (but with a relatively short {\sl off} time) or, more likely, from temporarily disruption of the jet due to jet-ISM interaction (or a combination of the two) resulting in a kind of {\sl ``sputtering''} emission that would keep the lobes fed, albeit perhaps at low rate. In particular, the second scenario would also result in a strong decollimation of the jets, slowing down the flow in the lobes and originating their low surface brightness. This scenario would also better explain the presence of turbulence in the plasma flow.


\subsection{Relevance and connection with other sources}
\label{sec:relevance}

The results on NGC~3998 are relevant in connection with the more general effort to understand the nature and evolution of low power radio sources (including e.g. core galaxies, the so-called coreG \citealt{Balmaverde06,Baldi09}; galactic scale jets (GSJ) e.g. Webster in prep.; FR0 radio galaxies \citealt{Baldi19}, Capetti et al. subm.). 
What we have observed in NGC~3998 confirms one of the possible scenarios put forward to explain the lack of large, extended structures in these sources and, in particular, in FR0 galaxies: slower jets can be more subject to instabilities and entrainment, which causes their premature disruption and fading  (\citealt{baldi2015}). 
Thus, our results on NGC~3998 appear to support the scenario in which the lack of extended emission in these low-power sources would be due to the rapid fading of extended lobes. This would be due to a combination of interaction, temporarily disrupting  the plasma flow and the large entrainment of thermal gas.

Also among extended radio sources and radio galaxies there are cases which present a number of similarities with NGC~3998. Some examples have been discussed in \cite{Brienza18}. In particular it is worth mentioning B2~0258+35 and Cen~A as two cases of low-power radio galaxies with a very high surface brightness contrast between the central regions and the large scales lobes. 
In both cases, and similarly to NGC~3998, the large lobes, despite the low-surface brightness, have the spectral index of actively fuelled structures (i.e. $\alpha \sim 0.7$). Also in these two objects, the spectral index is quite homogeneous across the lobes with no trend that could be connected with the ageing of the electrons (see \citealt{Brienza18} and \citealt{Morganti99,Kinley18} for B2~0258+35 and Centaurus~A respectively). In B2~0258+35 the spectral index of the lobes does not show any significant curvature up to 6.6~GHz, suggesting even tighter requirements for the reacceleration of the electrons in the lobes.

In Centaurus~A, a low-surface brightness, large-scale jet connecting the bright central part with the large-scale lobes has been found \citep{Morganti99,Kinley18}. This structure would ensure that the supplies of energy from the central galaxy to the lobes is still occurring (even if at low rate as suggested by their low surface brightness). This fuelling process is likely also maintaining the level of  turbulence necessary for the in-situ particle reacceleration within the lobes, process that would prevent the steepening of the radio spectrum.

Other objects with radio properties (and radio power) similar to NGC~3998 are the Seyfert galaxies Mrk~6 and NGC~6764 \citep{Kharb06,Kharb10}. Particularly interesting is the fact that also in these objects the diffuse lobes/bubble have a standard  homogeneous spectral index (see Mrk~6, \citealt{Kharb06}) and  S-shaped morphology, there explained as precessing jets, have been also observed. An interaction temporarily disrupting the jet is also suggested (see NGC~6764, \citealt{Kharb10}). 

In summary, the view we are getting from NGC~3998 and  other similar objects is telling us that amorphous, low surface brightness structures should not be taken, by default, as emission from remnant structures in low-power radio AGN.
Furthermore, the proposed (``sputtering'') scenario has consequences for the morphology of the radio source. It would influence the way the source (and in particular the large scale emission) evolves and the way the source is classified. 

Indeed, although objects like NGC~3998 seem to be rare, it may be worth mentioning that the number of known low-power radio AGN with similar properties (albeit with a radio power more in the range of FRI radio galaxies) is now growing.
In particular, the core dominance  and the low surface brightness of the lobes (a few mJy/arcmin$^2$, i.e. at the limit of what detectable with present day radio telescopes) have been proposed to identify radio sources with intermittent activity (see e.g. \citealt{Saripalli12,Kuzmicz17,Brienza18}, Jurlin et al. in prep.).
 For example, about 15\% of radio sources found in the LOFAR image of the Lockman Hole area \citep{Brienza17} are found to have a high core dominance (e.g. $S_{core}/S_{tot}> 0.2$) and an amorphous, low-surface brightness extended emission (see Jurlin et al. in prep.). Some of these sources have a radio structures reminiscent the one of NGC~3998.  Thus, they form not only an interesting but also a relevant group of sources (see also \citealt{Mingo19} for a detailed discussion of the classification). The combination of the LOFAR images with images at 1400~MHz becoming available, for common areas, from the APERTIF phased-array feed surveys on the Westerbork Synthesis Radio Telescope, will allow to obtain the necessary spectral index information.
 

\section{Conclusions}
\label{sec:conclusions}

In this paper we have presented new LOFAR 147~MHz observations of the radio source associated with the galaxy NGC~3998.
The new LOFAR observations have confirmed the properties of the radio structure observed at 1400~MHz, i.e. the sources is core-dominated with low-surface brightness lobes. Most importantly,
the observations have further shown that the lobes do not have an ultra-steep spectrum as expected if they would represent old remnant structures. Instead they have a spectral index typical of still active structures (with an average value of $\alpha^{\rm 1400 MHz}_{\rm 147 MHz} \sim 0.6$). Furthermore, the spectral index is relative homogeneous across the lobes. 

From the spectral properties we have estimated a (upper) limit to the age of the lobes (38~Myr) while conservative assumptions about the propagation of the jets suggests ages of the lobes between a few times $10^6$ to at most $10^7$ yr.
The results suggest the scenario where the lobes are still actively fed  by  the  central  region  (or  where  the  injection  is  only very recently stopped).
Although the possibility of an intermittent fuelling (with short {\sl off} time) is supported by the variability of the core of NGC~3998 and may be likely occurring, the high contrast between the central regions and the large-scale lobes, and the spectral properties, suggest the plasma to be turbulent and in-situ reacceleration to occur in the lobes.  This can happen if the jet is temporarily disrupted by the interaction with a dense and rich circumnuclear medium.

Our findings on NGC~3998 would suggest that the lack of extended emission in low-power radio sources (e.g. FR0, core galaxies, galactic scale jets) could be due to the rapid fading of extended lobes. This would originate by a combination of interaction, temporarily disrupting the plasma flow and the large entrainment of thermal gas.
Furthermore, the view we are getting from NGC~3998 and the other similar objects is telling us that amorphous, low surface brightness structure should not be taken, by default, as emission from remnant structures. 

If the proposed scenario is correct, tracing the evolution of such low power radio sources would be made more difficult by the relatively short time scale of their extended lobes if they quickly fade to very low surface-brightness structure.
However, a growing group of radio sources with properties similar to NGC~3998 is being identified in the deep (and high spatial resolution) surveys carried out at low frequencies by LOFAR (i.e. the LOFAR Two-metre Sky Survey, LoTTS (\citealt{Shimwell19}).
Thus, these sources (see e.g. \citealt{Mingo19}, Jurlin in prep.)
can tell us about evolutionary paths of the radio plasma and/or about specific conditions of the ISM different from what found in classical radio galaxies - i.e.  FRI and II - helping completing our knowledge about the how radio emission shapes and is shaped by the surrounding medium.  

\begin{acknowledgements}
We thank the anonymous referee for their helpful comments. The research leading to these results has received funding from the European Research Council under the European Union’s Seventh Framework Programme (FP/2007-2013)/ERC Advanced Grant RADIOLIFE-320745. LOFAR, the Low Frequency Array designed and constructed by ASTRON, has facilities in several countries, that are owned by various parties (each with their own funding sources), and that are collectively operated by the International LOFAR Telescope (ILT) foundation under a joint scientific policy. 
\end{acknowledgements}

\bibliographystyle{aa}
\bibliography{Paper_ref.bib}

\end{document}